\documentclass{article}
\usepackage{ijcai20}

\usepackage{times}

\usepackage{soul}
\usepackage{url}
\usepackage[hidelinks]{hyperref}
\usepackage[utf8]{inputenc}
\usepackage[small]{caption}
\usepackage{graphicx}
\usepackage{amsmath}
\usepackage{booktabs}
\usepackage{algorithm}
\usepackage{algorithmic}
\usepackage{amssymb}
\usepackage{xcolor}
\usepackage{comment}

\newcommand{\setalglineno}[1]{%
	\setcounter{ALC@line}{\numexpr#1-1}}

\title{Anomaly Detection on Seasonal Metrics via Robust Time Series Decomposition}

\author{
Tianwei Li$^{1,2}$\footnote{Tianwei Li is currently a PhD candidate at Rutgers University. This research was done when Tianwei worked as a full-time intern at eBay.}\and
Yitong Geng$^2$\footnote{Author to whom any correspondence should be addressed.}\and
Huai Jiang$^{2}$
\affiliations
$^1$Rutgers University, NJ\\
$^2$eBay Inc.\\
\emails
tianwei.li@rutgers.edu,
\{yigeng, huajiang\}@ebay.com
}

\begin{document}

\maketitle

\begin{abstract}
The stability and persistence of web services are important to Internet companies to improve user experience and business performances. 
To keep eyes on numerous metrics and report abnormal situations, time series anomaly detection methods are developed and applied by various departments in companies and institutions.
In this paper, we proposed a robust anomaly detection algorithm (MEDIFF) to monitor online business metrics in real time. 
Specifically, a decomposition method using robust statistical metric--median--of the time series was applied to decouple the trend and seasonal components. 
With the effects of daylight saving time (DST) shift and holidays, corresponding components were decomposed from the time series. 
The residual after decomposition was tested by a generalized statistics method to detect outliers in the time series. 
We compared the proposed MEDIFF algorithm with two open source algorithms (SH-ESD and DONUT) by using our labeled internal business metrics. 
The results demonstrated the effectiveness of the proposed MEDIFF algorithm.
\end{abstract}

\section{Introduction}
To ensure stabilization or increase in revenue, technology companies especially Internet companies providing online streaming services to monitor their business metrics and solve metric related troubles in real time. 
Anomaly detection is applied to discover unexpected breakouts caused by external factors such as vicious Internet attacks or internal factors such as services breakdowns. 
Corresponding actions such as restoration and maintenance will be took immediately based on the alarms.

Monitoring and detecting anomalies in multiple business key performance indicators (KPIs) in real time is a challenging problem in industrial. 
First, the number of anomalies in the millions of real-time business KPIs time series is much smaller than that of non-anomalies. 
The detection algorithm is required to accurately capture the anomalies from huge amount of metrics. 
Second, no standard feature of anomalies leads to the lack of labels. 
As a result, supervised learning algorithms \cite{laptev2015generic}\cite{liu2015opprentice}\cite{shipmon2017time}, in general, are not suitable for industrial application. 
Moreover, it is inevitable that missing or/and error values exist in the business metrics time series under service maintenance. 
Thus, the anomaly detection algorithm should be robust to these missing or/and error values.
Finally, to achieve monitor business metrics in real time, the response of the anomaly detection algorithm should be fast, such as less than 20$\%$ of metrics sampling period. 
Existing deep learning algorithms \cite{zhang2019time}\cite{xu2018unsupervised} may not be suitable for this requirement as the training process is time consuming.
Therefore, a fast and robust anomaly detection method is required to be developed. 

In this paper, a statistical based robust anomaly detection algorithm, MEDIFF, was proposed to monitor the online business metrics in real time at eBay. 
Our goal was to detect and report both short term deviations, e.g., large spikes during several minutes, and long term deviations, e.g., a skewing during an hour caused by unstable Internet service provided by telecommunication companies, in the metrics with one-minute sampling period in a short response time.
The MEDIFF technique was developed to meet the requirement of robust and fast response detection on streaming business metrics. 
The proposed method was based on a robust statistical trend-seasonal decomposition model and the generalized extreme Studentized deviate (ESD) many-outlier detection technique \cite{rosner1983percentage}. 
Specifically, the time series was decomposed into the trend and the seasonal components with a moving median smoothing and a short window week-over-week median values, respectively. 
Then, the residual was computed by removing the trend and the seasonal components from the time series.
Finally, the outliers were detected by applying the generalized ESD test on the residual. 
As applying the robust median metric and the statistical analysis, the response of MEDIFF was both fast and unimpaired by anomalies that happened in history. 
The proposed MEDIFF algorithm was evaluated by implementing experiments on our labeled internal business production dataset. 

The pattern of the time series that we considered was sensitive to the effect of daylight saving time (DST) shift and holidays due to the small sampling period. 
We analyzed the performance of MEDIFF on the time series during the DST and holiday periods and proposed a solution to compensate the effect of DST and holidays. 
Specifically, the seasonal trend component during DST and the effect component during holidays were captured by moving median smoothing, respectively. 
Then, the DST seasonal was obtained by combining the week-over-week seasonal component with the seasonal trend component under chosen weights.
After removing these components (trend, DST seasonality, and holiday effect) from the time series, the residual was tested by ESD to detect the outliers. 
The performance of the DST/holidays compensation method was also evaluated by our internal business metrics. 

The rest of the paper is organized as follows. 
In section \ref{sec_related_work}, previous works related to time series anomaly detection were discussed. 
In section \ref{sec_mediff_esd}, the proposed robust anomaly detection algorithm MEDIFF and the compensation method for DST and holidays were presented in details. 
The experiment implementation and results were described and evaluated in section \ref{sec_experiment}.
Conclusions were commented in section \ref{sec_conclusion}.

\section{Related Works}
\label{sec_related_work}
Current anomaly detection techniques used by large Internet companies are basically developed in three ways based on respective requirements. 
The statistical SH-ESD algorithm developed by Twitter \cite{vallis2014novel,hochenbaum2017automatic} works for their cloud business metrics with strong seasonality. 
The technique employs STL decomposition \cite{cleveland1990stl} to determine the seasonal component of a given time series and then applies extreme Studentized deviate (ESD) \cite{rosner1975detection} on the residual to detect the anomalies. 
The STL decomposition \cite{cleveland1990stl} requires at least seven season periods to capture the accurate seasonal component. 
However, this requirement results in huge amount of data for the time series with one minute sampling period and weekly seasonality in our application at eBay. 
Fbprophet developed by Facebook \cite{taylor2018forecasting} was proposed to solve challenges associated with large variety of time series. 
This method can be used to solve anomaly detection problems by fitting a regression model with interpretable parameters and predicting the future behaviors.

Supervised learning methods such as EGADS \cite{laptev2015generic} and Opprentice \cite{liu2015opprentice} take advantage of machine learning technique to improve the detection accuracy. 
Anomaly detectors are trained by using user feedbacks as labels and anomaly scores as features. 
However, the small percentage of anomalies in time series leads to unbalanced classes in the training dataset. 
Furthermore, the labels provided by users might be with low accuracy and therefore aggravates the unbalance. 
SR-CNN model proposed by Microsoft is a combination of unsupervised algorithm with supervised learning model \cite{ren2019time}. 
Spectral residual provides high accuracy labels to the convolutional neural networks to further improve the output accuracy of the anomaly detection. 

State-of-the-art unsupervised learning models, such as Gaussian Mixture Model (GMM) \cite{laxhammar2009anomaly}, Support Vector Machine (SVM) based classifier \cite{erfani2016high}, Long Short Term Memory networks (LSTM) \cite{malhotra2016lstm}, Variational autoencoders (VAE) \cite{zhang2019time} and its extension DONUT \cite{xu2018unsupervised}, have been developed in recent years to overcome the challenge of inaccuracy or lack of anomaly labels. 
These models basically focus on learning normal patterns rather than learning abnormal patterns whenever possible. 
The anomalies detected by these methods are generally based on the built-in anomaly scores, resulting in uninterpretable outliers.

\section{Proposed Technique}
\label{sec_mediff_esd}
\begin{figure}[t]
	\centering
	\includegraphics[scale=0.23]{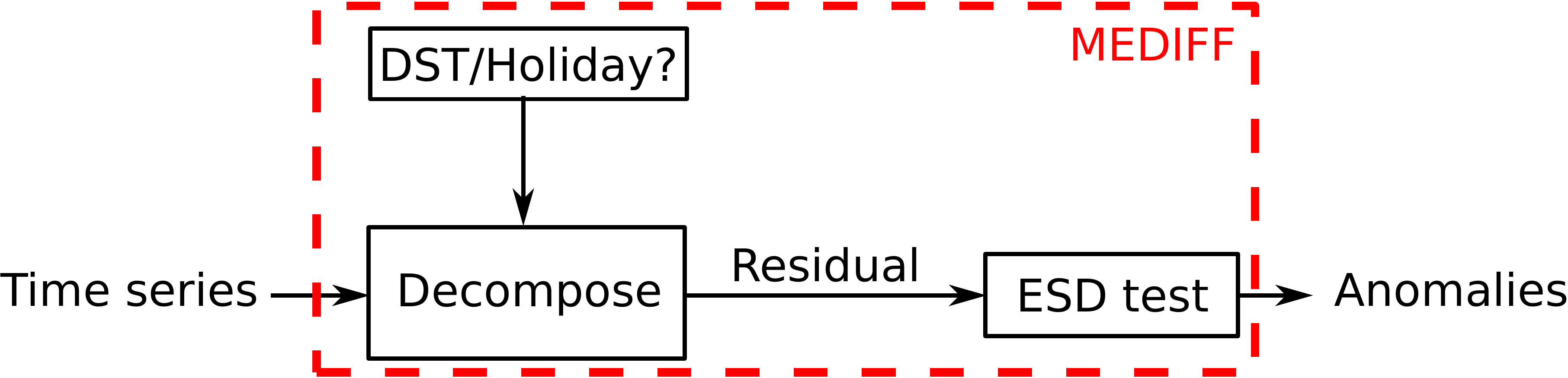}
	\caption{The schematic representation of the MEDIFF detector: 
		Time series is first decomposed to extract trend component, seasonal component, and DST shift or holiday effect components (if applicable). 
		Then, the residual is tested by ESD technique to detect the anomalies (outliers).}
	\label{fig_steps}
\end{figure}

In this section, the proposed robust anomaly detection technique, MEDIFF, is presented in detail. 
The algorithm is mainly divided into the decomposition phase and the test phase as shown in Fig. \ref{fig_steps}. 
In general, the time series data is decomposed to extract the trend and the seasonal components. 
The components of DST and holiday effects, if applicable, are also extracted from the time series. 
Next, the residual obtained by removing the extracted components (trend, seasonality, DST and holiday effects) from the time series is tested by the generalized extreme Studentized deviate (ESD) many-outlier detection technique \cite{rosner1983percentage} to detect the anomalies in the time series.

\subsection{Time Series Decomposition}
\label{sec_mediff}
In general, time series data can be decomposed into three components \cite{cleveland1990stl}\cite{kawasaki2006structural}, i.e.,
\begin{equation}
\label{eq_decompose3}
y=\mu+s+r
\end{equation}
where $y$, $\mu$, $s$, and $r$ denote the time series raw data, the trend component, the seasonal component, and the residual, respectively. 
To decrease the skewing of outliers and avoid the effect of missing or error data in time series, here, we propose the MEDIFF decomposition to capture the components in Eq. (\ref{eq_decompose3}) by using the robust statistical metric---median. 
Specifically, given a time series $y[k]$ with length $\ell_y$ (i.e., $k\in \mathbb{N}$ (set of natural numbers) and $1 \leqslant k \leqslant \ell_y$, denotes as $k \in \mathbb{N}[1,\ell_y]$ in the rest of the paper), the trend component is decomposed by using moving median technique with the window length of $w_\mu$, i.e.,
\begin{equation}
\label{eq_trend}
\mu[k]=\mathbf{median}\Big(\bigcup_{i=0}^{w_\mu-1} y[k-i]\Big), \quad \mbox{for}~k \in \mathbb{N}[w_\mu, \ell_y]
\end{equation}
where the window length $w_\mu$ is chosen in practice as the length of one season $\ell_s$ of the time series. 
As the moving median in Eq. (\ref{eq_trend}) uses history data with length $w_\mu$, the length of the trend component is $\ell_y-w_\mu+1$, i.e., the first $w_\mu-1$ elements are truncated. 
Then, the seasonal component is decomposed by first removing the trend component $\mu[k]$ from the time series $y[k]$, i.e., the detrended time series $\hat{y}[k]$
\begin{equation}
\label{eq_detrend}
\hat{y}[k]=y[k]-\mu[k], \qquad \mbox{for}~k \in \mathbb{N}[w_\mu,\ell_y]
\end{equation}
and then evaluating the median of the data at the same window area position $[-w_s, w_s]$ in each season, i.e.,
\begin{equation}
\label{eq_seasonal}
\begin{split}
s[k]&=\mathbf{median}\Big(\bigcup_{i=0}^{n_s-1}\bigcup_{j=-w_s}^{w_s}\hat{y}[k \pm i \ell_s+j]\Big),\\
&~\mbox{for}~k\in\mathbb{N}[w_\mu,\ell_y],~\mbox{and}~(k \pm i \ell_s+j) \in \mathbb{N}[w_\mu,\ell_y]
\end{split}
\end{equation}
where $n_s$ and $\ell_s$ denote the number of seasons in the detrended time series $\hat{y}[k]$ and the length of each season, respectively. 
The residual $r[k]$ is obtained by removing the trend and seasonal components from the time series, i.e.,
\begin{equation}
\label{eq_residual}
r[k]=y[k]-\mu[k]-s[k], \qquad \mbox{for}~k \in \mathbb{N} [w_\mu,\ell_y]
\end{equation}

The MEDIFF decomposition is fast and computational saving compared to the STL decomposition \cite{cleveland1990stl} as no need to compute the local weight for each data \cite{cleveland1979robust} in STL. 
Moreover, MEDIFF decomposition is robust to both single outliers and short term skewing. 
Specifically, as the seasonal component is unchangeable overtime (i.e., seasonal component in each season is the same), the short term skewing happened in only one season (due to network attacks or unexpected short term sales promotion) is not considered as part of the seasonal component, therefore, will be detected as anomalies.

\begin{figure}[t]
	\centering
	\includegraphics[scale=0.4]{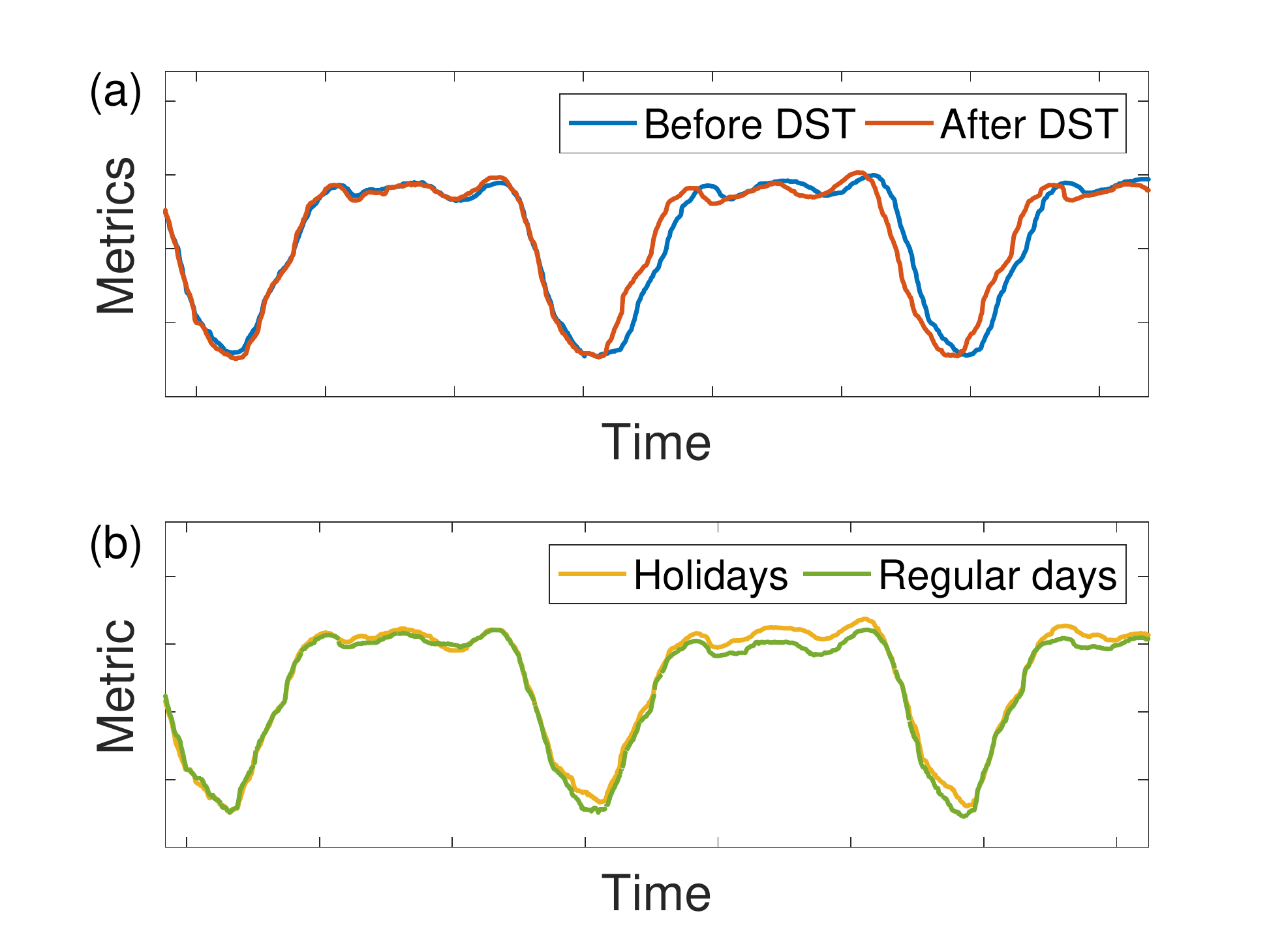}
	\caption{The effect of DST (a) and holidays (b) on time series metrics, respectively.}
	\label{fig_dst_holiday}
\end{figure}

\subsection{Decomposition of Daylight Saving Time Shift and Holidays Components}
\label{sec_dst}
The time series related to customer behaviors such as logging in and checking out is heavily affected by daylight saving time (DST) shift and holidays \cite{kamstra2000losing}\cite{mcelroy2018modeling}. 
Specifically, customers keep their daily activities according to their local time regardless of the DST shift, i.e., customer activities happen at the similar time of each day/week before or after local DST shift. 
However, the time series recorded on the server present different (lagging or advancing) patterns as the server time does not shift with DST. 
Thus, the time series in one season (e.g., one week) after DST is not overlapped with that before DST (see Fig. \ref{fig_dst_holiday}(a)). 
Such a changing pattern caused by DST shift results in inaccurate seasonal component by MEDIFF. 
Similar effect by holidays exists. 
Specifically, customer activities may burst as promotions provided by merchants during the Black Friday or reduce as people are away from Internet to enjoy vacations during Christmas and New Year (see Fig. \ref{fig_dst_holiday}(b)).

Due to the effects of DST and holidays, the decomposition components (trend and seasonality) introduced in Subsec. \ref{sec_mediff} cannot represent the feature of the time series. 
We propose to include components in the decomposition to compensate the effects of DST and holidays \cite{taylor2018forecasting}, respectively. 
Specifically, the time series is decomposed as following
\begin{equation}
\label{eq_decompose4}
y=\mu+s^D+\gamma e+\epsilon
\end{equation}
where $\gamma \in \{0,1\}$ is a user defined constant ($\gamma=1$ when considering the effect of holidays/events in the time series, otherwise $\gamma=0$), and $s^D$, $e$ and $\epsilon$ denote the seasonal component during DST effect period (called as ``{\em DST seasonal component}" in the rest of the paper), the effects of events (holidays, etc.) and the residual which is considered as Gaussian noise, respectively. 
For the time series $y[k]$ with length $\ell_y$, the trend component is first obtained as in Eq. (\ref{eq_trend}). 
Then, the DST seasonal component $s^D[k]$ is obtained as 
\begin{equation}
\label{eq_season_DST}
s^D[k]=\beta s[k] + (1-\beta) \hat{s}[k]
\end{equation}
where $\beta \in [0,1]$ is a user defined weight to balance the DST effect, $s[k]$ is the seasonal component obtained by Eq. (\ref{eq_seasonal}), and $\hat{s}[k]$ denotes the seasonal trend obtained by using moving median with a small window length $\hat{w}_s$, i.e., 
\begin{equation}
\label{eq_seasonal_trend}
\begin{split}
\hat{s}[k]&=\mathbf{median}\Big(\bigcup_{i=0}^{\hat{w}_s-1} \hat{y}[k-i]\Big),\\
&~\mbox{for}~k\in \mathbb{N}[w_\mu, \ell_y],~\mbox{and}~(k-i) \in \mathbb{N}[w_\mu,\ell_y]
\end{split}
\end{equation}
After removing the trend and the DST seasonal components from the time series, the event component is decomposed by using moving median with the window length of $w_r$, i.e.,
\begin{equation}
\label{eq_event}
\begin{split}
r[k]&=y[k]-\mu[k]-s^D[k],\\
&\quad \mbox{for}~k\in \mathbb{N}[w_\mu,\ell_y],~\mbox{then}\\
e[k]&=\mathbf{median}\Big(\bigcup_{i=0}^{w_r-1}r[k-i]\Big),\\
&\quad \mbox{for}~k\in \mathbb{N}[w_\mu+w_r-1,\ell_y]
\end{split}
\end{equation}
Finally, the decomposed residual is computed as 
\begin{equation}
\label{eq_epsilon}
\epsilon[k]=r[k]-\gamma e[k], \quad \mbox{with}~\gamma \in \{0,1\}
\end{equation}

In general, the time series decomposition by MEDIFF can be presented as Eq. (\ref{eq_decompose4})--Eq. (\ref{eq_epsilon}) with $\gamma=0$ and $\beta=1$ when the effects of holidays/events and DST are near to negligible level in the time series $y$, i.e., during normal days/weeks with no DST or holidays/events. 
The dates/time of DST shift and holidays are saved in the sever to indicate the detector to apply appropriate parameters $\gamma$ and $\beta$.
Specifically, $\gamma =1$ or $\beta$ will be defined by the user ($\beta \in [0,1)$) when time horizon of the time series $y$ contains the date of holiday/event or the DST shift date/time, respectively. 

\subsection{Outliers Detection by ESD Test}

After the MEDIFF decomposition, the decomposed residual $\epsilon[k]$ for $k\in \mathbb{N}[w_\mu+w_r-1,\ell_y]$ will be checked by the generalized ESD test \cite{rosner1983percentage} to detect the unspecified number of outliers. 
For completeness, we briefly describe below the implementation of the generalized ESD test in the proposed MEDIFF detector.

Given the upper bound on the suspected number of outliers $m$, the generalized ESD test first computes corresponding statistic $z$-score of the decomposed residual $\epsilon$ for $m$ iterations by removing the observation that maximizes the $z$-score from the decomposed residual in each iteration, i.e.,
\begin{equation}
\label{eq_esd_zscore} 
z_i=\mathbf{max}\frac{|\epsilon_i-\bar{\epsilon}_i|}{\sigma_i},
\qquad i=1,2,\cdots, m
\end{equation}
where $\epsilon_i$, $\bar{\epsilon}_i$ and $\sigma_i$ denote the remaining sample series of the decomposed residual $\epsilon$ in the $i^{th}$ iteration, the mean and the standard deviation of $\epsilon_i$ in the $i^{th}$ iteration, respectively. 
To further avoid the skewing effect of the outliers, we modified Eq. (\ref{eq_esd_zscore}) by replacing the mean $\bar{\epsilon}_i$ and the standard deviation $\sigma_i$ with the median $\tilde{\epsilon}_i$ and median absolute deviation (MAD) $\tilde{\sigma}_i$ of $\epsilon_i$, i.e.,
\begin{equation}
	\label{eq_esd_zscore_mad}
	\tilde{z}_i=\mathbf{max}\frac{|\epsilon_i-\tilde{\epsilon}_i|}{\tilde{\sigma}_i},
	\qquad i=1,2,\cdots, m
\end{equation}
where $\tilde{\epsilon}_i=\mathbf{median}(\epsilon_i)$ and $\tilde{\sigma}_i=\mathbf{median}\left(|\epsilon_i-\tilde{\epsilon}_i|\right)$. 
Then, the ESD computes the critical value for $m$ times
\begin{equation}
\label{eq_esd_critical}
\begin{split}
\lambda_i=\frac{(n-i)t_{p,n-i-1}}{\sqrt{(n-i-1+t_{p,n-i-1}^2)(n-i+1)}},\\ \qquad i=1,2,\cdots, m
\end{split}
\end{equation}
where $t_{p,\nu}$ is the $100p$ percentage point from the $t$-distribution with $\nu$ degrees of freedom and
\begin{equation}
p=1-\frac{\alpha}{2(n-i+1)}
\end{equation}
with $\alpha$ the significance level.
The number of anomalies is determined by finding the largest $i$ such that $\tilde{z}_i>\lambda_i$.
Then, the anomalies are determined at the corresponding observations with statistic $z$-score index less than the number of the anomalies. In general, $\tilde{z}_i$ may not be always larger than the critical value $\lambda_i$ before permanently dropping below it.

The above MEDIFF algorithm (decomposition and the ESD test) is summarized in Algorithm~\ref{alg:mediff}.

\begin{algorithm}[h]
	\caption{MEDIFF detector}
	\label{alg:mediff}
	\textbf{Input}: Time series $y[k]$\\
	\textbf{Parameter}: $w_\mu$, $w_s$, $\hat{w}_s$, $w_r$, $\beta$, $\gamma$\\
	\textbf{Output}: Anomalies in the time series $y[k]$\\
	Decomposition:
	\begin{algorithmic}[1] 
		\STATE Extract the trend component $\mu[k]$ by Eq. (\ref{eq_trend});
		\STATE Remove $\mu[k]$ from $y[k]$ by Eq. (\ref{eq_detrend});
		\STATE Extract the seasonal component $s[k]$ and the seasonal trend component $\hat{s}[k]$ by Eq. (\ref{eq_seasonal}) and Eq. (\ref{eq_seasonal_trend}), respectively;
		\IF {DST effect exists}
		\STATE Obtain the DST seasonal component $s^D[k]$ by Eq. (\ref{eq_season_DST}) with $\beta \in [0,1)$;
		\ELSE
		\STATE Obtain $s^D[k]$ by Eq. (\ref{eq_season_DST}) with $\beta=1$;
		\ENDIF
		\STATE Remove $\mu[k]$ and $s^D[k]$ from $y[k]$ and extract the event component $e[k]$ by Eq. (\ref{eq_event});
		\IF {event/holiday effect exists}
		\STATE Obtain the residual $\epsilon[k]$ by Eq. (\ref{eq_epsilon}) with $\gamma=1$;
		\ELSE
		\STATE Obtain $\epsilon[k]$ by Eq. (\ref{eq_epsilon}) with $\gamma =0$.
		\ENDIF
	\end{algorithmic}
	ESD test:
	\begin{algorithmic}[1]
		\setalglineno{15}
		\STATE Specify the suspected number of outliers $m$;
		\WHILE{$i \leqslant m$} 
		\STATE Calculate the $z$-score $\tilde{z}_i$ of the decomposed residual $\epsilon[k]$ by Eq. (\ref{eq_esd_zscore_mad}) and remove the maximum observation from $\epsilon[k]$;
		\ENDWHILE
		\STATE Calculate the critical values $\lambda_i$ by Eq. (\ref{eq_esd_critical});
		\STATE Compare $\tilde{z}_i$ and $\lambda_i$, and determine the outliers in $\epsilon[k]$;
		\STATE \textbf{return} the outliers as the anomalies in $y[k]$.
	\end{algorithmic}
\end{algorithm}

\section{Experiments and Discussion} 
\label{sec_experiment}
In this section, we presented and discussed the performance of the proposed MEDIFF algorithm, compared with two open source anomaly detection algorithms---SH-ESD \cite{vallis2014novel,hochenbaum2017automatic} and DONUT \cite{xu2018unsupervised}.

\subsection{Experiment Implementation}
We applied two groups of internal business metrics during different months (group $A$ from September to the end of October and group $B$ from November to the end of December) with five pieces of metrics in each group. 
The metrics had strong weekly seasonality, i.e., the week-over-week seasonal components were nearly overlapped. 
Each piece of metrics contained about 80,640 data points with time interval of one minute, i.e., one data point per minute for about 60 days. 
The anomalies in each metric were labeled manually by expert operators. 
Any successive anomalies (skewing) were considered as one anomaly and denoted by labeling the first one of these anomalies. 
By visualizing the metrics in each group, no effect of DST or holidays was found on the metrics in group $A$, while both DST (in November) and holidays (Black Friday and Christmas) effected on the metrics in group $B$. 
Those dates of DST and holidays were recorded in the system for MEDIFF decomposition on the metrics in group $B$. 

\begin{table*}[t]
	\centering
	\caption{Performance on metrics in group $A$ (w/o DST and holiday)}
	\begin{tabular}{ c c c c c }
		\hline
		Algorithm & Precision & Recall & F1-score & Running time (s) \\
		\hline
		MEDIFF ($\gamma=0,\beta=1$) & 0.5361 & 0.8500 & 0.6575 & 0.9754\\		
		SH-ESD & 0.6215 & 0.4128 & 0.4961 & 11.4971\\		
		DONUT & 0.5572 & 0.875 & 0.6808 & 189.0562 (training) + 0.0678\\
		\hline
	\end{tabular}
	\label{tab_A}
\end{table*}

\begin{table*}[t]
	\centering
	\caption{Performance on metrics in group $B$ (with DST or holiday)}
	\begin{tabular}{ c c c c c }
		\hline
		Algorithm & Precision & Recall & F1-score & Running time (s) \\
		\hline
		MEDIFF ($\gamma=0,\beta=1$) & 0.5952 & 0.4832 & 0.5334 & 0.9562\\
		MEDIFF ($\gamma=1,\beta=0.4$) & 0.775 & 0.6603 & 0.7131 & 0.9847\\		
		SH-ESD & 0.5774 & 0.3846 & 0.4617 & 11.9975\\
		DONUT & 0.5386 & 0.6240 & 0.5782 & 185.7478 (training) + 0.0737\\
		\hline
	\end{tabular}
	\label{tab_B}
\end{table*}

In the experiment, we implemented the MEDIFF detection on time series batches with length 40,320 (i.e., 4 weeks) as the application in our online real-time production environment.
Specifically, we partitioned each piece of metrics in both group $A$ and $B$ into two batches with 40,320 data points in each batch. 
Then, the time series in each batch in group $A$ was decomposed and the corresponding residual was obtained by following the Algorithm \ref{alg:mediff} with fine-tuned parameters $w_\mu=10080$, $w_s=3$, $\hat{w}_s=30$, and $w_r=60$, and weight $\gamma=0$ and $\beta=1$, while for the batches in group $B$, the same window length $w_\mu$, $w_s$, $\hat{w}_s$, and $w_r$ were applied and the weight were chosen as $\gamma=1$ and $\beta=0.4$. 
After the decomposition, the residual of each batch was tested by ESD with suspected number of outliers $m \approx 600$ (i.e., anomaly rate is $0.02$) and the significance level $\alpha=0.05$.

To evaluate the performance of MEDIFF, open source anomaly detection algorithms SH-ESD \cite{vallis2014novel,hochenbaum2017automatic} and DONUT \cite{xu2018unsupervised} 
were also tested on the labeled metrics in both group $A$ and $B$. 
For SH-ESD, the algorithm was also applied on each time series batch of length 4 weeks. 
For DONUT, we trained the models for each piece of metrics with the time series in the first 4 weeks (September) in group $A$ and then tested the metrics in the next 4 weeks (October). 
To test the performance of DONUT during DST and holiday periods, the model was first trained on the time series in October (i.e., metrics in the last 4 weeks in group $A$) and then applied to detect anomalies in the two batches in group $B$ (November and December), respectively.
We tried our best to achieve good performance of the algorithms on the experiment dataset.

\subsection{Experiment Results and Discussion}
We evaluated the performance of the algorithms by calculating the commonly used machine learning metrics {\em Precision} ($P$), {\em Recall} ($R$), and {\em F1-score} ($F_1$), i.e., 
\begin{equation}
\begin{split}
P=\frac{\mbox{TP}}{\mbox{TP}+\mbox{FP}}, \quad
R=\frac{\mbox{TP}}{\mbox{TP}+\mbox{FN}}, \quad
F_1=\frac{2 R P}{R + P}
\end{split}
\end{equation}
where TP, FP, and FN denote the number of positive values that are predicted correctly, the number of negative values that are wrongly predicted as positive, and the number of positive values that are wrongly predicted as negative, respectively. 
Similar as the special labeling method in the experiment dataset, a set of successive anomalies detected by all these three algorithms were condensed to one detected positive and denoted as the first one in this set. 
This kind of positive (condensed from successive anomalies) was considered as a true positive if the detected positive was within 10 minutes delay of the labeled positive in the experiment dataset, otherwise a false positive.

\begin{figure}[t]
	\centering
	\includegraphics[scale=0.4]{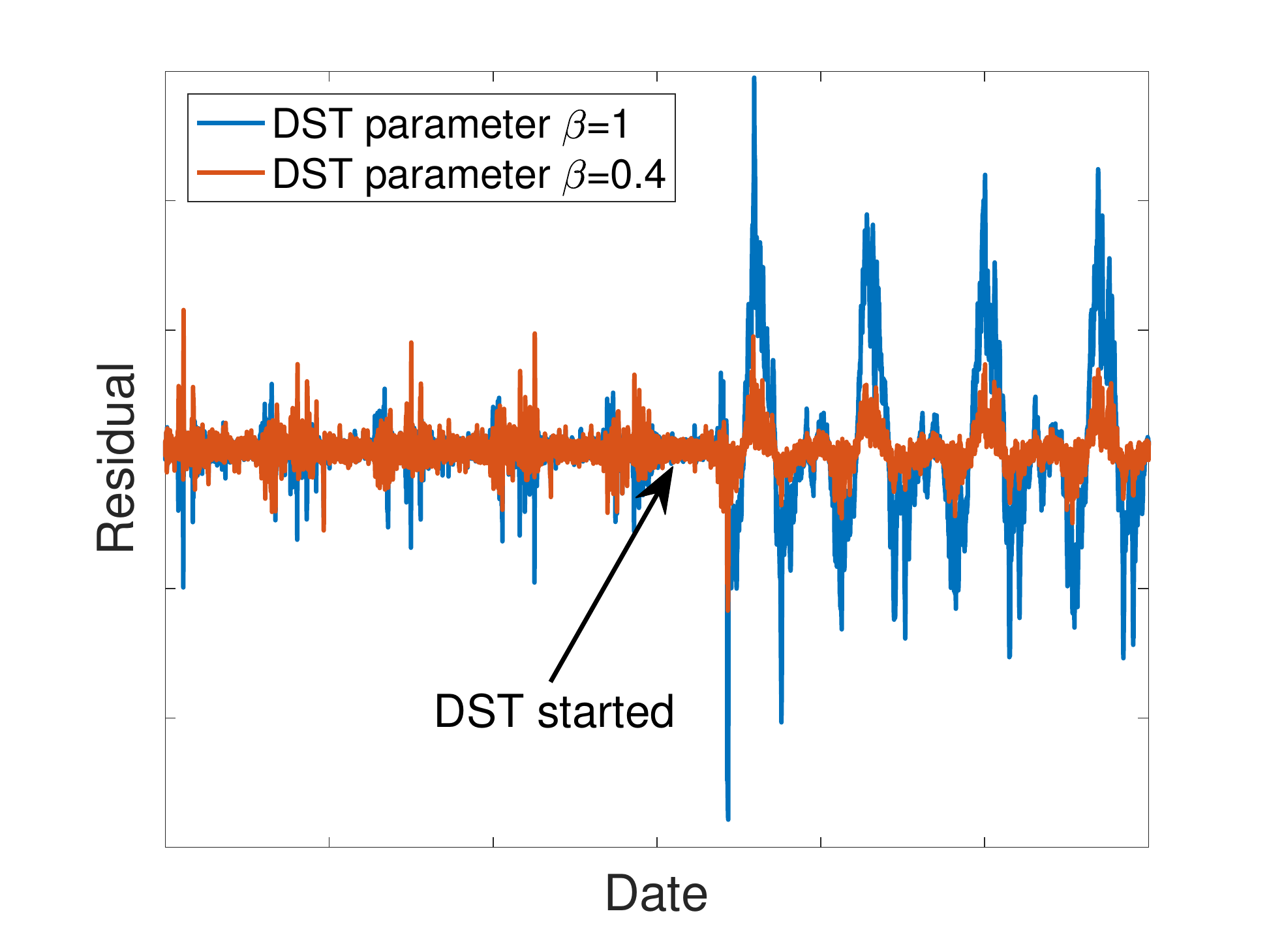}
	\caption{The residuals after decomposing a piece of time series with DST effect by $\beta=1$ (blue) and $\beta=0.4$ (red), respectively.}
	\label{fig_dst_result}
\end{figure}

The detection results of each batch metrics in both group $A$ and $B$ by all three algorithms were averaged and presented in Table \ref{tab_A} and Table \ref{tab_B}, respectively. 
Overall, the proposed MEDIFF was more effective (faster and better performance) on our internal seasonal metrics than other methods. 
First, MEDIFF detected more true positive anomalies than SH-ESD (higher recall value) on the metrics in both group $A$ and $B$. 
Such an advantage was due to the accurate robust MEDIFF decomposition, i.e., most of the prominent outliers in the decomposed residuals were true positives. 
However, by the STL decomposition in SH-ESD, the residual might be contaminated by the decomposed components (trend, seasonality), therefore, true positives might be left out. 
Then, the performance of both MEDIFF and SH-ESD declined on the metrics effected by DST (group $B$) as neither of the decomposition methods was suitable. 
The results were improved by applying the MEDIFF decomposition for DST and holiday period with parameters $\gamma=1$ and $\beta=0.4$. 
As the seasonal pattern shifted by DST was captured by the MEDIFF decomposition with $\gamma=1$ and $\beta=0.4$, the residual was clear from distortion of the seasonal component (see Fig. \ref{fig_dst_result}). 
Thus, the ESD test captured more true positives and reported less wrong anomalies, resulting in increased precision and recall values.
Finally, the DONUT algorithm had higher overall performance on the metrics in both groups compared to the MEDIFF (with $\gamma=0$ and $\beta=1$) and the SH-ESD. 
However, it may still be effected by DST or holiday as we found several unexpected and unexplainable anomalies on the metrics in group $B$, resulting in decreased precision value.

\section{Conclusion} 
\label{sec_conclusion}
We proposed a robust anomaly detection algorithm (MEDIFF) to monitor online business metrics in real time at eBay. 
A robust decomposition method using the moving median of the time series was first applied to decouple the trend, the seasonal, and the event components (if applicable). 
The effects of DST and holidays on the business metrics were analyzed and corresponding decomposition method was also handled by the proposed MEDIFF.
The residual after removing the decomposed components was tested by the generalized statistics test algorithm (ESD) to detect anomalies. 
The proposed MEDIFF algorithm was tested on our internal business metrics, compared with the SH-ESD and DONUT methods. 
The results were evaluated to demonstrate the proposed MEDIFF algorithm. 

\bibliographystyle{named}
\bibliography{ijcai20}

\begin{thebibliography}{}

\bibitem[\protect\citeauthoryear{Cleveland \bgroup \em et al.\egroup
  }{1990}]{cleveland1990stl}
Robert~B Cleveland, William~S Cleveland, Jean~E McRae, and Irma Terpenning.
\newblock Stl: a seasonal-trend decomposition.
\newblock {\em Journal of official statistics}, 6(1):3--73, 1990.

\bibitem[\protect\citeauthoryear{Cleveland}{1979}]{cleveland1979robust}
William~S Cleveland.
\newblock Robust locally weighted regression and smoothing scatterplots.
\newblock {\em Journal of the American statistical association},
  74(368):829--836, 1979.

\bibitem[\protect\citeauthoryear{Erfani \bgroup \em et al.\egroup
  }{2016}]{erfani2016high}
Sarah~M Erfani, Sutharshan Rajasegarar, Shanika Karunasekera, and Christopher
  Leckie.
\newblock High-dimensional and large-scale anomaly detection using a linear
  one-class svm with deep learning.
\newblock {\em Pattern Recognition}, 58:121--134, 2016.

\bibitem[\protect\citeauthoryear{Hochenbaum \bgroup \em et al.\egroup
  }{2017}]{hochenbaum2017automatic}
Jordan Hochenbaum, Owen~S Vallis, and Arun Kejariwal.
\newblock Automatic anomaly detection in the cloud via statistical learning.
\newblock {\em arXiv preprint arXiv:1704.07706}, 2017.

\bibitem[\protect\citeauthoryear{Kamstra \bgroup \em et al.\egroup
  }{2000}]{kamstra2000losing}
Mark~J Kamstra, Lisa~A Kramer, and Maurice~D Levi.
\newblock Losing sleep at the market: The daylight saving anomaly.
\newblock {\em American Economic Review}, 90(4):1005--1011, 2000.

\bibitem[\protect\citeauthoryear{Kawasaki}{2006}]{kawasaki2006structural}
Yoshinori Kawasaki.
\newblock A structural time series model facilitating flexible seasonality.
\newblock In {\em Proceedings of the Conference on Seasonality, Seasonal
  Adjustment and Their Implications for Short-term Analysis and Forecasting}.
  Citeseer, 2006.

\bibitem[\protect\citeauthoryear{Laptev \bgroup \em et al.\egroup
  }{2015}]{laptev2015generic}
Nikolay Laptev, Saeed Amizadeh, and Ian Flint.
\newblock Generic and scalable framework for automated time-series anomaly
  detection.
\newblock In {\em Proceedings of the 21th ACM SIGKDD International Conference
  on Knowledge Discovery and Data Mining}, pages 1939--1947. ACM, 2015.

\bibitem[\protect\citeauthoryear{Laxhammar \bgroup \em et al.\egroup
  }{2009}]{laxhammar2009anomaly}
Rikard Laxhammar, Goran Falkman, and Egils Sviestins.
\newblock Anomaly detection in sea traffic-a comparison of the gaussian mixture
  model and the kernel density estimator.
\newblock In {\em 2009 12th International Conference on Information Fusion},
  pages 756--763. IEEE, 2009.

\bibitem[\protect\citeauthoryear{Liu \bgroup \em et al.\egroup
  }{2015}]{liu2015opprentice}
Dapeng Liu, Youjian Zhao, Haowen Xu, Yongqian Sun, Dan Pei, Jiao Luo, Xiaowei
  Jing, and Mei Feng.
\newblock Opprentice: Towards practical and automatic anomaly detection through
  machine learning.
\newblock In {\em Proceedings of the 2015 Internet Measurement Conference},
  pages 211--224. ACM, 2015.

\bibitem[\protect\citeauthoryear{Malhotra \bgroup \em et al.\egroup
  }{2016}]{malhotra2016lstm}
Pankaj Malhotra, Anusha Ramakrishnan, Gaurangi Anand, Lovekesh Vig, Puneet
  Agarwal, and Gautam Shroff.
\newblock Lstm-based encoder-decoder for multi-sensor anomaly detection.
\newblock {\em arXiv preprint arXiv:1607.00148}, 2016.

\bibitem[\protect\citeauthoryear{McElroy \bgroup \em et al.\egroup
  }{2018}]{mcelroy2018modeling}
Tucker~S McElroy, Brian~C Monsell, and Rebecca~J Hutchinson.
\newblock Modeling of holiday effects and seasonality in daily time series.
\newblock {\em Statistics}, 1, 2018.

\bibitem[\protect\citeauthoryear{Ren \bgroup \em et al.\egroup
  }{2019}]{ren2019time}
Hansheng Ren, Bixiong Xu, Yujing Wang, Chao Yi, Congrui Huang, Xiaoyu Kou, Tony
  Xing, Mao Yang, Jie Tong, and Qi~Zhang.
\newblock Time-series anomaly detection service at microsoft.
\newblock {\em arXiv preprint arXiv:1906.03821}, 2019.

\bibitem[\protect\citeauthoryear{Rosner}{1975}]{rosner1975detection}
Bernard Rosner.
\newblock On the detection of many outliers.
\newblock {\em Technometrics}, 17(2):221--227, 1975.

\bibitem[\protect\citeauthoryear{Rosner}{1983}]{rosner1983percentage}
Bernard Rosner.
\newblock Percentage points for a generalized esd many-outlier procedure.
\newblock {\em Technometrics}, 25(2):165--172, 1983.

\bibitem[\protect\citeauthoryear{Shipmon \bgroup \em et al.\egroup
  }{2017}]{shipmon2017time}
Dominique~T Shipmon, Jason~M Gurevitch, Paolo~M Piselli, and Stephen~T Edwards.
\newblock Time series anomaly detection; detection of anomalous drops with
  limited features and sparse examples in noisy highly periodic data.
\newblock {\em arXiv preprint arXiv:1708.03665}, 2017.

\bibitem[\protect\citeauthoryear{Taylor and
  Letham}{2018}]{taylor2018forecasting}
Sean~J Taylor and Benjamin Letham.
\newblock Forecasting at scale.
\newblock {\em The American Statistician}, 72(1):37--45, 2018.

\bibitem[\protect\citeauthoryear{Vallis \bgroup \em et al.\egroup
  }{2014}]{vallis2014novel}
Owen Vallis, Jordan Hochenbaum, and Arun Kejariwal.
\newblock A novel technique for long-term anomaly detection in the cloud.
\newblock In {\em 6th $\{$USENIX$\}$ Workshop on Hot Topics in Cloud Computing
  (HotCloud 14)}, 2014.

\bibitem[\protect\citeauthoryear{Xu \bgroup \em et al.\egroup
  }{2018}]{xu2018unsupervised}
Haowen Xu, Wenxiao Chen, Nengwen Zhao, Zeyan Li, Jiahao Bu, Zhihan Li, Ying
  Liu, Youjian Zhao, Dan Pei, Yang Feng, et~al.
\newblock Unsupervised anomaly detection via variational auto-encoder for
  seasonal kpis in web applications.
\newblock In {\em Proceedings of the 2018 World Wide Web Conference}, pages
  187--196. International World Wide Web Conferences Steering Committee, 2018.

\bibitem[\protect\citeauthoryear{Zhang and Chen}{2019}]{zhang2019time}
Chunkai Zhang and Yingyang Chen.
\newblock Time series anomaly detection with variational autoencoders.
\newblock {\em arXiv preprint arXiv:1907.01702}, 2019.

\end{thebibliography}

\end{document}